\documentclass[12pt]{article}

\usepackage{hyperref}
\usepackage{times}
\usepackage{natbib}
\usepackage{graphicx}

\title{Paleomagnetic evidence for a disk substructure in the early solar system}

\author
{Cau\^e S. Borlina$^{1\ast}$, Benjamin P. Weiss$^1$, James F. J. Bryson$^2$,\\ Xue-Ning Bai$^3$, Eduardo A. Lima$^1$,
Nilanjan Chatterjee$^1$, Elias N. Mansbach$^1$\\
\\
\normalsize{$^1$Department of Earth, Atmospheric and Planetary Sciences,}\\
\normalsize{Massachusetts Institute of Technology, Cambridge, MA, USA. }\\

\normalsize{$^2$Department of Earth Sciences, Oxford
University, Oxford, UK.}\\ 

\normalsize{$^3$Institute for Advanced Study and Department of Astronomy,} \\
\normalsize{Tsinghua University, Beijing, China}.\\
\\
\normalsize{$^\ast$Corresponding author. Email: \href{mailto:caue@mit.edu}{caue@mit.edu}}\\
\\
\normalsize{Published in \textit{Science Advances}: \href{https://www.science.org/doi/10.1126/sciadv.abj6928}{DOI:10.1126/sciadv.abj6928}}
}

\date{}

\begin{document} 


\baselineskip16pt


\maketitle


\begin{abstract}
 Astronomical observations and isotopic measurements of meteorites suggest that substructures are common in protoplanetary disks and may even have existed in the solar nebula. Here, we conduct paleomagnetic measurements of chondrules in CO carbonaceous chondrites to investigate the existence and nature of these disk sub-structures. We show that the paleomagnetism of chondrules in CO carbonaceous chondrites indicates the presence of a 101 $\pm$ 48 $\mu$T field in the solar nebula in the outer solar system ($\sim$3 to 7 AU from the Sun). The high intensity of this field relative to that inferred from inner solar system ($<$3 AU) meteorites indicates a factor of $\sim$5 to 150 mismatch in nebular accretion between the two reservoirs. This suggests substantial mass loss from the disk associated with a major disk substructure, possibly due to a magnetized disk wind.
\end{abstract}

\section{Introduction}
Observations from the Atacama Large Millimeter/submillimeter Array have shown that substructures, mostly in the form of rings and gaps, are prevalent in protoplanetary disks \citep{andrews_observations_2020}. Isotopic studies of meteorites and their components have been interpreted as evidence that two isotopically distinct regions existed within $\sim$7 astronomical units (AU) from the young Sun (see the Supplementary Materials), known as the noncarbonaceous ($<$3 AU) and carbonaceous reservoirs (3 to 7 AU) \citep{kruijer_great_2020, scott_isotopic_2018}, that existed during the protoplanetary disk phase of the solar system (i.e., solar nebula). It has been proposed that these two reservoirs were separated by a gap in the disk, perhaps generated by the growth of Jupiter \citep{kleine_non-carbonaceouscarbonaceous_2020} and/or a pressure local maximum in the disk \citep{brasser_partitioning_2020}. Alternatively, these two reservoirs may have formed because of a migrating snowline with no persistent disk gap \citep{lichtenberg_bifurcation_2021}. Because the evolution of protoplanetary disks is thought to depend on the coupling of the weakly ionized gas of the disk with large-scale magnetic fields \citep{bai_heat_2009, bai_global_2017, turner_transport_2014, weiss_history_2021}, we can search for evidence of disk substructures and explore their origin by studying the paleomagnetism of meteorites that formed in each reservoir.

Previous paleomagnetic measurements of LL chondrites, derived from the noncarbonaceous reservoir, indicate the presence of a disk midplane magnetic field of 54 $\pm$ 21 $\mu$T at 2.0 $\pm$ 0.8 million years (Ma) after the formation of calcium-aluminum–rich inclusion (CAIs) \citep{fu_solar_2014,kita_evolution_2011}. These paleointensities, which were measured from individual chondrules that carry thermoremanent magnetization (TRM) acquired before their accretion onto the LL parent body, provide evidence for the existence of a nebular magnetic field in the noncarbonaceous reservoir. Paleomagnetic studies of CM \citep{cournede_early_2015}, CR \citep{fu_weak_2020}, and CV \citep{fu_fine-scale_2021} chondrites indicate a field in the carbonaceous reservoir of $>$6 $\mu$T at $\sim$3 Ma after CAI formation, $<$8 $\mu$T at $\sim$3.7 Ma after CAI formation \citep{weiss_history_2021, schrader_distribution_2017}, and $\geq$40 $\mu$T sometime between $\sim$3-40 Ma after CAI formation \citep{fu_fine-scale_2021, carporzen_magnetic_2011}. However, these records have several key limitations. For instance, the records in CM and CV chondrites are postaccretional chemical remanent magnetizations acquired during parent-body alteration \citep{cournede_early_2015}. This poses two problems: The magnetic record could have been imparted by a parent-body dynamo field rather than from the solar nebula field and the retrieved paleointensity is likely a lower limit \citep{weiss_history_2021}. In addition, the age of the CR record is within error of the estimated lifetime of nebula \citep{wang_lifetime_2017}, such that it may not constrain the strength of the nebular field during the main period of disk accretion \citep{weiss_history_2021}. Therefore, the intensity of the nebular field in the carbonaceous reservoir is currently poorly constrained. 

To obtain robust paleointensity records from the midplane of the solar nebula in the carbonaceous reservoir, we conducted paleomagnetic studies on two CO carbonaceous chondrites: Allan Hills (ALH) A77307 (type 3.03) and Dominion Range (DOM) 08006 (type 3.00) \citep{bonal_organic_2007, davidson_mineralogy_2019, alexander_mutli-technique_2018, scott_shock_1992, alexander_origin_2007}. We selected these meteorites because they experienced low peak metamorphic temperatures (200$^{\circ}$ to 300$^{\circ}$C), minor parent-body aqueous alteration, shock pressures below 5 GPa, and minimal terrestrial weathering (weathering grades A/B and Ae, respectively) \citep{bonal_organic_2007, davidson_mineralogy_2019, alexander_mutli-technique_2018, scott_shock_1992, alexander_origin_2007}. Therefore, they are unlikely to have been magnetically overprinted following accretion onto the CO parent body, with DOM 08006 in particular being one of the least altered known meteorites \citep{davidson_mineralogy_2019}. Following the previous paleomagnetic study of LL chondrules, we targeted dusty olivine chondrules because they contain high-fidelity paleomagnetic recorders in the form of fine-grained ($\sim$25 to 1000 nm) kamacite ($\alpha$-Fe) crystals formed before accretion onto the parent body \citep{fu_solar_2014, shah_oldest_2018, einsle_multi-scale_2016, lappe_mineral_2011, lappe_comparison_2013}. Because chondrules cooled quickly in the protoplanetary disk environment [100$^{\circ}$ to 1000$^{\circ}$C hour$^{-1}$; \citep{scott_chondrites_2013}], they should carry a near-instantaneous TRM record of the nebular field \citep{fu_solar_2014, desch_importance_2012, desch_critical_2010}. Al-Mg dating of CO chondrules indicate that this record was acquired 2.2 $\pm$ 0.8 Ma after CAI formation \citep{kita_evolution_2011}\footnote{Recent high-precision Al-Mg ages of LL chondrules support a shorter formation interval
than previous Al-Mg ages. See Supplementary Text for more information.}. 

We extracted six 100- to 300-$\mu$m-diameter dusty olivine chondrules from both meteorites: two from ALHA77307 (DOC1 and DOC2) and four from DOM 08006 (DOC3, DOC4, DOC5, and DOC6). Three of the DOM 08006 chondrules were split into two subsamples each (DOC3a, DOC3b, DOC5a, DOC5b, and DOC6a, DOC6b) to produce nine total subsamples from both meteorites used for paleomagnetic measurements. All chondrules and chondrule fragments were mutually oriented during extraction and paleomagnetic measurements. 

Given the chondrules’ weak natural remanent magnetization (NRM) (ranging from 1.3$\times$10$^{-10}$ down to 1.7$\times$10$^{-12}$ Am$^2$ before demagnetization), we obtained magnetic measurements using the superconducting quantum interference device (SQUID) microscope and quantum diamond microscope (QDM) in the Massachusetts Institute of Technology (MIT) Paleomagnetism Laboratory (see Materials and Methods) \citep{weiss_paleomagnetic_2007}.

\section{Results}
Backscattered electron microscopy (BSE) images and compositional analysis using wavelength dispersive spectrometry (WDS) indicate that the chondrules contain numerous submicrometer diameter inclusions of nearly pure-Fe kamacite (see the Supplementary Materials). Furthermore, QDM imaging confirms that the magnetization-carrying capacity of the chondrules is dominated by these grains rather than by any multidomain metal grains and/or secondary ferromag- netic minerals (see the Supplementary Materials). These fine metal grains are expected to have formed during subsolidus reduction of the chondrules before their accretion on the parent body \citep{lappe_mineral_2011}. On the basis of their size and composition, many of these grains are predicted to be in the single vortex size range and smaller, which has been shown to have the potential to carry paleomagnetic records over a period longer than the lifetime of the solar system \citep{fu_solar_2014, nagy_nano_2019}.

\begin{figure}[ht!]
\centering
\includegraphics[scale=0.5]{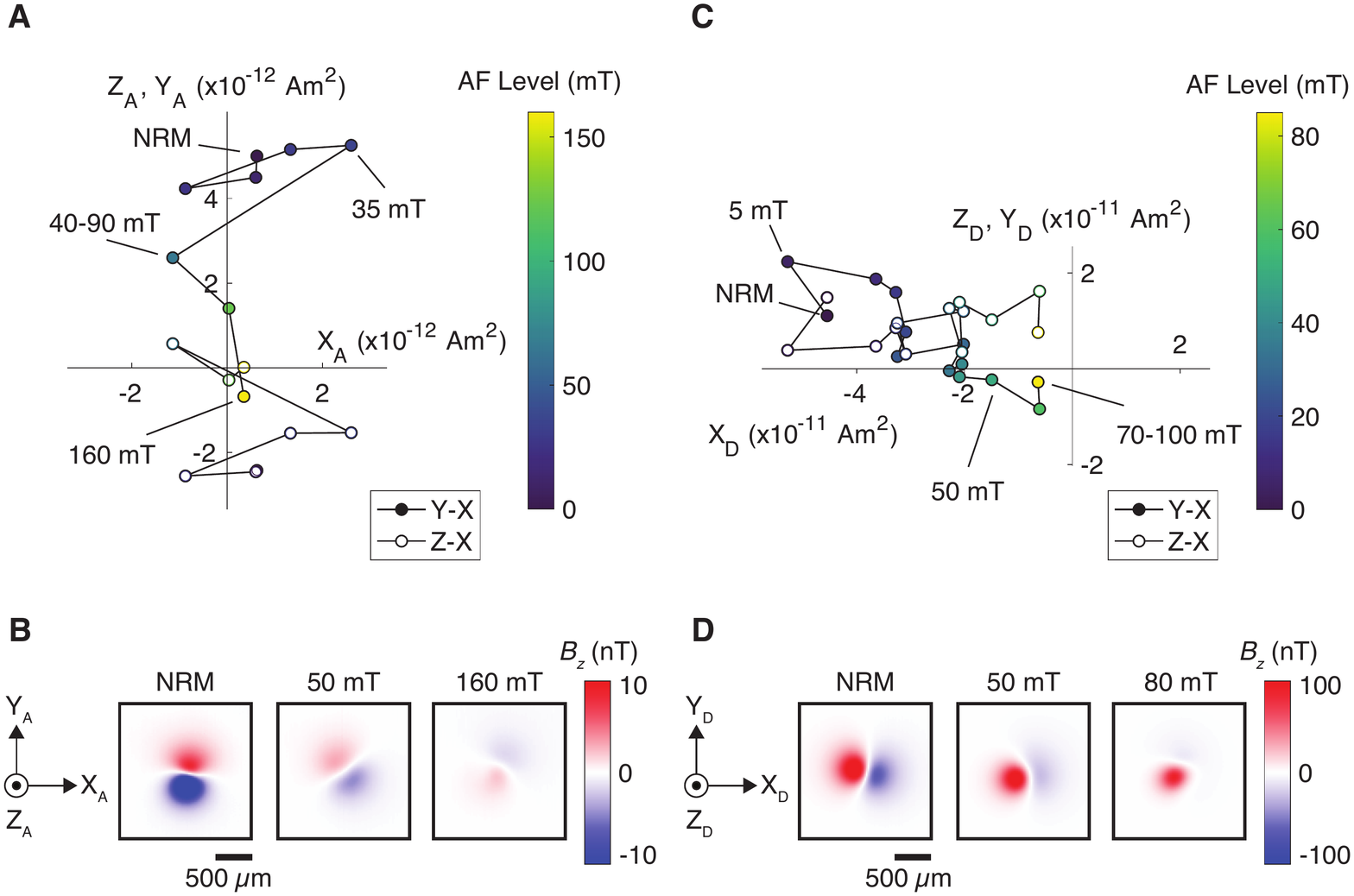}
\caption{\textbf{AF demagnetization of CO dusty olivine chondrules.} (\textbf{A} and \textbf{B}) DOC1 from ALHA77307. (\textbf{C} and \textbf{D}) DOC4 from DOM 08006. (A and C) Orthographic projections
of NRM vector endpoints during alternating field (AF) demagnetization showing averaged measurements for repeated AF steps and across AF levels. Closed symbols
show the $Y$-$X$ projection of the magnetic moment, and open symbols show $Z$-$X$ projection of the magnetic moment; subscripts “A” and “D” denote separate coordinate systems for ALHA77307 and DOM 08006, respectively. We interpret the steps between NRM and 160 mT for DOC1 and between 50 and 850 mT for DOC4 as constituting the HC components. Selected demagnetization steps are labeled. Color scales show the AF levels. (B and D) Out-of-the-page magnetic field component ($B_z$) maps for selected steps measured at a height of $\sim$300 $\mu$m above the chondrules obtained with the SQUID microscope. Each map represents one of six maps associated with different applications of the AF field to obtain each step shown in the orthographic projection.}
\label{fig:zijd}
\end{figure}

Our alternating field (AF) demagnetization showed that some subsamples carried a low-coercivity (LC) component blocked up to $<$20 mT (Fig. \ref{fig:zijd} and figs. S2 and S3). The LC component may be a viscous remanent magnetization acquired on Earth and/or a weak-field isothermal remanent magnetization acquired during sample handling. After the removal of the LC component, all subsamples were found to contain high-coercivity (HC) components blocked up to at least 50 mT (Fig. \ref{fig:zijd} , figs. S2 and S3, and table S1), with two subsamples having HC components blocked up to 160 and 270 mT (Fig. \ref{fig:zijd}  and fig. S2).

\begin{figure}[ht!]
\centering
\includegraphics[scale=0.7]{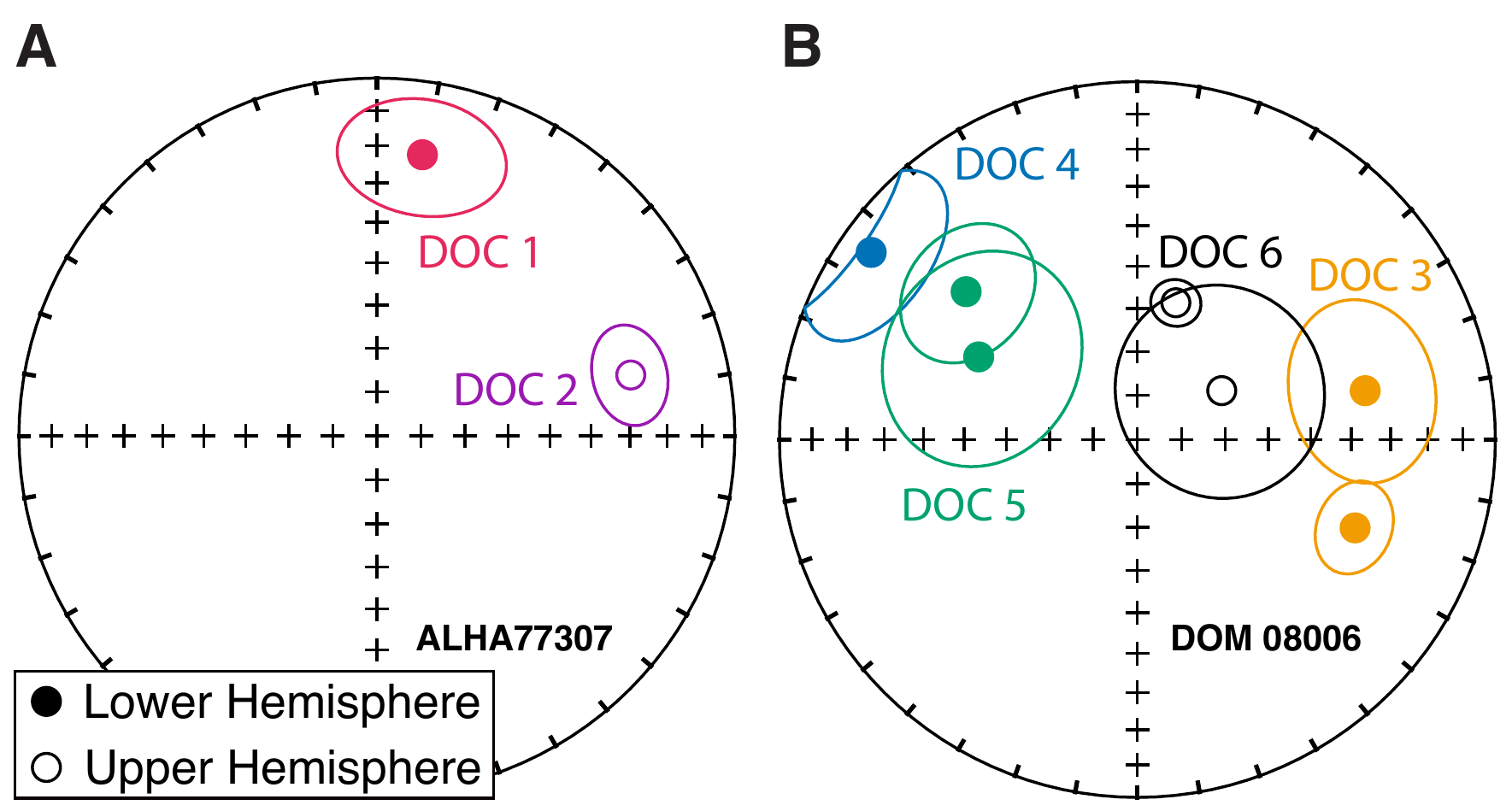}
\caption{\textbf{Direction of the HC components of the dusty olivine chondrules from
CO chondrites.} (\textbf{A}) ALHA77307. (\textbf{B}) DOM 08006. Shown are equal area stereographic
projections containing directions calculated from principal components analysis and their corresponding maximum angular deviations. Points with different
colors are from different individual chondrules, while points with identical colors are subsamples from an individual chondrule. The scattered directions from different chondrules in (A) and (B) indicate that the meteorites were not remagnetized since the chondrules accreted. The clustered directions among subsamples of the same chondrules in (B) are consistent with that expected for a TRM acquired during primary cooling in the solar nebula.}
\label{fig:equal}
\end{figure}

The high AF-stability of the HC components coupled with the pristine conditions of the meteorites suggest that the HC components are likely records of the nebular field. To further test this conclusion, we conducted unidirectionality tests and conglomerate tests (see the Supplementary Materials). Because the nebular field is expected to have been directionally homogeneous on submillimeter length scales, a nebular TRM should be unidirectional within each chondrule. Our measurements confirm this: Pairs of subsamples of three DOM 08006 chondrules have HC directions within each other’s maximum angles of deviation (Fig. \ref{fig:equal}). In addition, if the chondrules have not been remagnetized since parent body accretion (including on their parent body and after arrival on Earth), then they should collectively exhibit random magnetization directions. Our measure- ments of two chondrules from ALHA77307 and four chondrules from DOM 08006 (Fig. \ref{fig:equal}) confirm this: We cannot reject the hypothesis that both sets of directions are random with 95\% confidence \citep{watson_test_1956} (see the Supplementary Materials). In summary, the unidirectionality and conglomerate tests strongly support the conclusion that the chondrules carry robust paleomagnetic records of the solar nebula magnetic field acquired before accretion onto their parent bodies.

To determine the paleointensity of the recorded field, we compared the AF demagnetization of the NRM to that of an anhysteretic remanent magnetization (ARM) acquired in a bias field of 200 $\mu$T and a peak AF of up to 145 mT for seven chondrules [following previous studies \citep{tikoo_decline_2014}]. Paleointensity estimates were estimated assuming a ratio of ARM to TRM of 1.87 as previously measured for dusty olivine chondrules (see Materials and Methods). The resulting mean HC paleointensity estimates from two ALHA77307 chondrules and five DOM 08006 chondrules are 30 $\pm$ 10 $\mu$T and 59 $\pm$ 31 $\mu$T, respectively. Combining the seven samples and accounting for chondrule spinning during TRM acquisition [which decreases the background nebular field intensity recorded by the chondrule by an average factor of 2 \citep{fu_solar_2014}], we obtained a grand mean paleointensity of the background nebular field of 101 $\pm$ 48 $\mu$T (table S2).

\section{Discussion}
Together with the previous paleomagnetic study of CM chondrites \citep{cournede_early_2015}, the magnetic record from CO chondrules strongly supports the presence of a nebular magnetic field in the carbonaceous reservoir at $\sim$2 to 3 Ma after CAI formation. Furthermore, the data from the CO chondrules provide the first accurate constraints on the intensity of the nebular magnetic field in the carbonaceous reservoir. In particular, the CO chondrule paleointensities are $>$16 times higher than the lower limit measured from bulk CM chondrites, which highlights the importance of measuring TRMs to obtain robust magnetic records. The identification of magnetic fields in the noncarbonaceous and carbonaceous reservoirs suggests a widespread role for magnetically driven accretion in the early solar system.

\begin{figure}[ht!]
\centering
\includegraphics[scale=0.5]{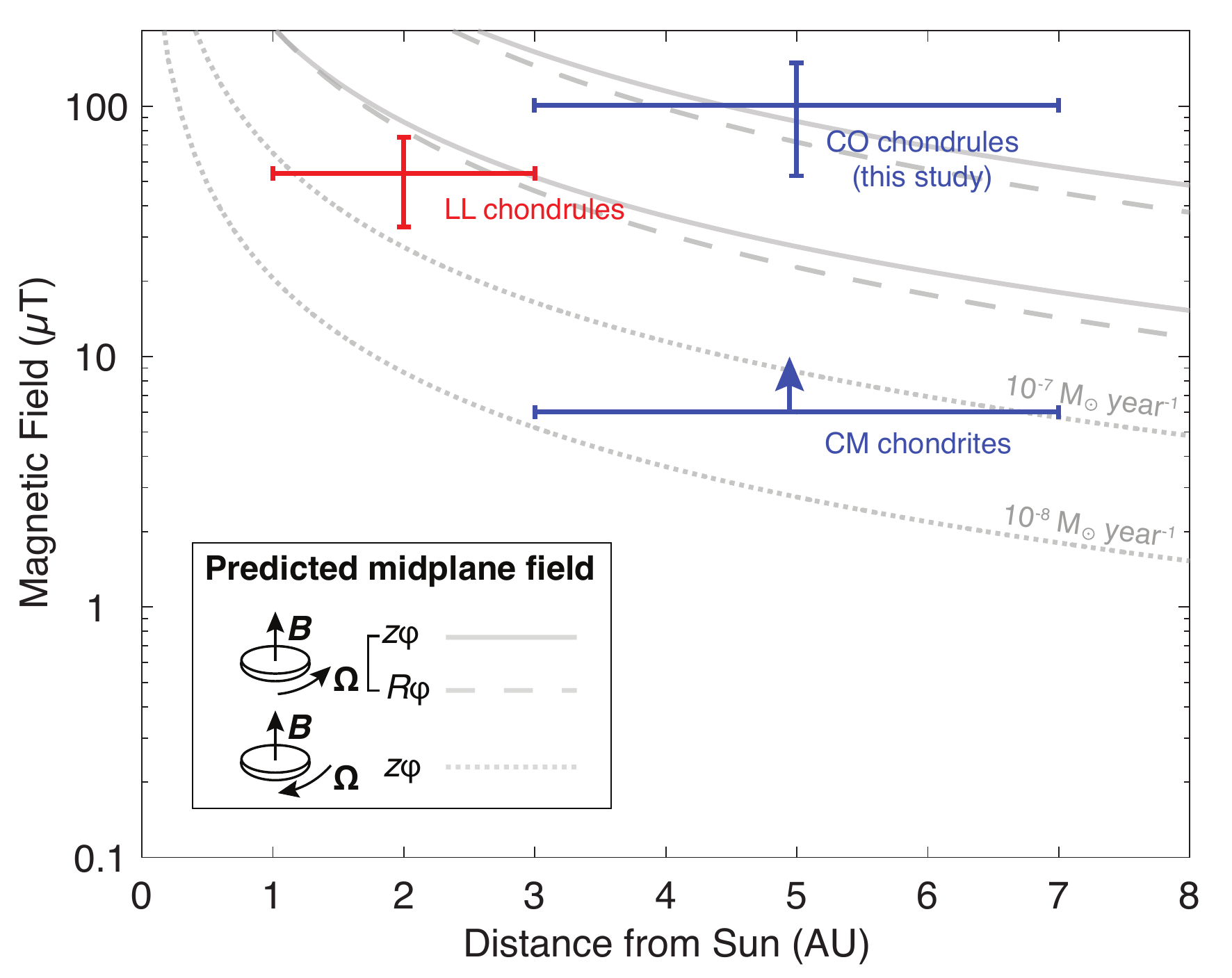}
\caption{\textbf{Comparison between paleomagnetic constraints and model predictions for the solar nebula magnetic field intensity.} Shown are paleointensity
records from the first 3 Ma after CAI formation: LL chondrules \citep{fu_solar_2014}, CM chondrites
\citep{cournede_early_2015}, and CO chondrules (this study). Solid and dashed lines show predicted midplane
magnetic field, due to vertical-toroidal [$z\varphi$; equation 3 of \cite{weiss_history_2021}] and radial-toroidal
[$R\varphi$; equation 2 of \cite{weiss_history_2021}] Maxwell stresses, respectively, assuming the nebular
magnetic field and sense of disk rotation are aligned. Dotted lines show the field
due to vertical-toroidal Maxwell stresses [$z\varphi$; equation 3 of \cite{weiss_history_2021}] assuming the
nebular magnetic field and sense of disk rotation are antialigned. Top and bottom curves were calculated assuming accretion rates of 10$^{-7}$ and 10$^{-8}$ $M_\odot$ year$^{-1}$ respectively.}
\label{fig:paleo}
\end{figure}

The structures and evolution of protoplanetary disks are governed by the mechanisms that drive disk accretion, likely mediated by magnetic fields. The accretion rate scales quadratically with field strength in the disk midplane with a prefactor depending on disk microphysics (especially ionization and field orientation). For a spatially constant accretion rate, the field intensity should decay as $\sim$R$^{-5/4}$ or $\sim$R$^{-11/8}$, where $R$ is the radial distance from the Sun, depending on whether accretion is primarily driven by the radial-toroidal
($R\varphi$) or vertical-toroidal ($z\varphi$) components of the Maxwell stress (Fig. \ref{fig:paleo}) \citep{weiss_history_2021}. Because of the Hall effect, the predicted field intensity is a factor of up to $\sim$10 higher if the background field threading the disk is aligned with disk rotation axis compared to the antialigned case. Given typical astronomically observed disk accretion rates of $\sim$1$\times$10$^{-8}$ $M_\odot$ year$^{-1}$ \citep{hartmann_accretion_1998}, the measured CO paleointensity strongly favors the scenario of aligned polarity (Fig. \ref{fig:paleo}), which would otherwise lead to an unreasonably large accretion rate ($\sim$1$\times$10$^{-5}$ $M_\odot$ year$^{-1}$). 

Considering the mean paleointensities from LL and CO chondrules, we find that the accretion rate was highly variable in the early solar system: for a disk with aligned polarity and a net vertical magnetic field ($z\varphi$ only), the accretion rates are $\sim4^{+4}_{-2} \times$10$^{-9}$ $M_\odot$ year$^{-1}$ in the noncarbonaceous reservoir and $\sim1^{+2}_{-0.6} \times$10$^{-7}$ $M_\odot$ year$^{-1}$ in the carbonaceous reservoir (Fig. \ref{fig:paleo}). The observed factor of $\sim$5 to 150 variation in the magnetically driven accretion rate between the two reservoirs could reflect variations in the accretion rate in time and/or in space. Temporal variations would be broadly consistent with astronomically observed accretion bursts in protoplanetary disks that occur on a timescale of hundreds of years \citep{hartmann_accretion_2016}. However,
our Monte Carlo simulations suggest that the probability that the observed changes in accretion are due to temporal fluctuations is $<$0.4\% (see the Supplementary Materials). Thus, our results favor the
presence of a spatial mismatch in the magnetically driven accretion rates between the two reservoirs. This spatial mismatch in the accretion rate has also been observed in recent paleomagnetic measurements of the CV chondrite Allende \citep{fu_fine-scale_2021}. While a spatial variation in accretion rate has also been proposed to explain the anomalously weak fields recorded by CR chondrules, it was not possible to conclusively differentiate between varying accretion rates or a null magnetic record because of the prior dissipation of the solar nebula in those samples \citep{weiss_history_2021}.

The observed mismatch in the accretion rates requires a mechanism that removed mass from the accretion flux between the carbonaceous and the noncarbonaceous reservoirs. If proto-Jupiter or another giant planet was present between the two reservoirs, then it is possible that part of this accretion flux was intercepted and accreted onto the growing planet. However, if the mismatch was completely due to accretion onto proto-Jupiter, that would require a growth time scale of just $\sim$10,000 years for the planet. This is several orders of magnitude faster than the several–million year time scale predicted by the core accretion model, the favored mechanism for Jupiter formation \citep{atreya_weiss_2018}. Alternatively, a large fraction of accretion mass flux could be lost through a disk outflow. A substantial mass loss has been predicted in theoretical models of photoevaporation leading to inside-out disk clearing [e.g., \citep{clarke_dispersal_2001, owen_theory_2012, picogna_dispersal_2019}]. On the other hand, given the likely role magnetic fields in driving disk accretion, photo-evaporation and magnetized disk winds may operate concurrently \citep{wang_global_2019, bai_toward_2016}, leading to magnetothermal disk winds whose mass loss rates are comparable to or exceed that of observed accretion rates of protoplanetary disks. This is sufficient to account for our observed mismatch in accretion rates and potentially lead to the formation of an inner cavity (i.e., extreme version of a gap) \citep{suzuki_evolution_2016}. Note, however, that this scenario does not preclude a role for Jupiter in gap formation. Its presence would likely accelerate the formation of the cavity \citep{rosotti_interplay_2013}, transforming the solar nebula into a transition disk.

In summary, we present a robust record of magnetic fields in the carbonaceous reservoir of the solar nebula. When compared to previous measurements of magnetic fields from the noncarbonaceous reservoir and models that describe the magnetic field in protoplanetary disks, we observe that the accretion rate in the carbonaceous reservoir is several times higher than that of the noncarbonaceous reservoir, implying the presence of a mismatch in accretion rates. This mismatch may be associated with disk mass loss through the presence of a gas giant, photoevaporation, and/or magnetized winds. These mechanisms could produce a disk substructure like those observed astronomically and like that inferred from the isotopic dichotomy measured among various meteorites in the early solar system. 

\section{Materials and Methods}

\subsection{Chondrule extraction and orientation}
Bulk samples of ALHA77307,157 (0.57 g) and DOM 08006,102 (2.8 g) were obtained from the NASA Johnson Space Center. At MIT, oriented thick sections were cut from these using a wire saw cooled with ethanol during cutting. Each thick section had an average area of 4 cm$^2$ and a thickness of $\sim$500 $\mu$m. The thick sections were then polished down to 1-$\mu$m roughness, and dusty olivine chondrules were identified using reflected light microscopy. Overall, we found that dusty olivine chondrules are very rare among CO chondrites, with a frequency of only $\sim$1 out of 100 chondrules ($\sim$0.005 inclusion mm$^{-3}$). The six dusty olivine chondrules used in this study were obtained from two thick sections from ALHA77307 and eight from DOM 08006. We note that some thick sections did not contain identifiable dusty olivine chondrules. All extracted chondrules were at least 3 mm away from the fusion crust of the parent sample (table S1).

Figure S1 shows the procedure for chondrule extraction. Before extraction, the thick section surface orientation was documented using imaging with a petrographic microscope (step 1, fig. S1). A region of $\sim$300 to 500 $\mu$m in thickness was excavated around the target chondrule using a degaussed carbide dental drill bit (step 2, fig. S1). After excavation, we used two different techniques to extract the chondrules. In the first technique, chondrule samples DOC1, DOC2, DOC3a, and DOC3b were extracted using a degaussed dental drill then glued to the quartz disk using cyanoacrylate cement. While the glue hardened, we oriented the sample by comparing its surface with the previously acquired image of the chondrule in situ. In the second technique, chondrule samples DOC4, DOC5a, DOC5b, DOC6a, and DOC6b each had a $\sim$400 $\mu$m by 400 $\mu$m by 1000 $\mu$m non-magnetic quartz glass coverslip glued with cyanoacrylate cement onto their top surfaces (step 3, fig. S1). We then marked the glass with a marker to orient it with respect to the thick section. The chondrules were then extracted using a degaussed dental drill bit and mounted on a quartz glass with cyanoacrylate cement. After extraction, we added Kapton tape as standoffs that were at least 2 mm away from the samples (step 4, fig. S1). The standoffs were added such that they were slightly higher than the samples, protecting the samples from rubbing against the SQUID microscope window during the magnetic measurements. Table S1 shows the orientations of the chondrules during the SQUID microscope measurements. We estimate that the extraction and mounting techniques can add up to 15$^{\circ}$ of total angular uncertainty.

\subsection{Paleomagnetism}
We used AFs to demagnetize the chondrules in steps of 5 or 10 mT, using an automatic 3-axis degausser system integrated into the 2G Enterprises Superconducting Rock Magnetometer 755R \citep{kirschvink_rapid_2008} at MIT. We demagnetized the samples with repeated AF applications to reduce spurious ARM and used the Zijderveld-Dunlop averaging method to correct for gyroremanent magnetization \citep{stephenson_three-axis_1993}. The maximum AF field necessary to demagnetize the samples varied among our samples. Sample DOC1 was demagnetized up to 400 mT, DOC2 up to 410 mT, DOC3a up to 100 mT, DOC3b up to 60 mT, DOC4 up to 100 mT, DOC5a up to 60 mT, DOC5b up to 75 mT, DOC6a up to 70 mT, and DOC6b up to 100 mT. For each AF step, we measured the magnetic field of each sample six times: once after applications of the AF in the $X$, $Y$, and $Z$ directions successively, twice after applications in the $X$ direction, twice after applications in the $Y$ direction, and once after an application in the $Z$ direction.

NRM measurements were obtained using the SQUID microscope in the MIT Paleomagnetism Laboratory \citep{weiss_paleomagnetic_2007}. Measurements of the samples’ magnetic fields were obtained at an effective sensor-to-sample distance of $\sim$300 $\mu$m \citep{weiss_paleomagnetic_2007}. For samples found to be dipolar magnetic field sources (DOC1, DOC2, DOC3a, DOC3b, DOC4, DOC5b, and DOC6b), we used a previously described inversion technique to obtain the magnetic moment from the magnetic field \citep{lima_ultra-high_2016}. For samples DOC5a and DOC6a, whose fields were found to be nondipolar, we upward-continued the magnetic maps by 150 $\mu$m and retrieved their dipole moments using averages from a first- to the fifth- and second-degree multipole model, respectively. After obtaining magnetic moments from each AF step, we averaged across the six steps and sometimes also across AF levels. The directions of NRM components were calculated using principal components analysis \citep{kirschvink_least-squares_1980}. The demagnetization projections are shown in figs. S2 and S3. If a NRM component had a deviation angle less than the maximum angle deviation, then this component was inferred to be the characteristic component and therefore anchored to the origin \citep{kirschvink_least-squares_1980, tauxe_strength_2004}. Table S1 shows the results of the principal components analysis including the levels that were averaged.

To obtain paleointensities, we used the ARM method \citep{fu_solar_2014, tikoo_decline_2014} for which:

\begin{equation}
B_{paleo} = \frac{B_{lab}}{f'}\frac{\Delta NRM}{\Delta ARM} 
\end{equation}

where $B_{paleo}$ is the ancient magnetic field recovered from the experiment, $B_{lab}$ is the 200-$\mu$T ARM bias magnetic field applied to the sample, $\Delta NRM$ and $\Delta ARM$ are the respective changes in magnetic moment during the demagnetization of the NRM and the ARM, and $f’$ is the ratio of TRM to ARM. The ARM was applied with a peak AF field of 145 mT for DOC1 and DOC2, 100 mT for DOC3a, 60 mT for DOC3b, 100 mT for DOC4, 75 mT for DOC5b, and 100 mT for DOC6b. We AF demagnetized the ARM using the same sequence used for the NRM. All samples used for paleointensity determination were nearly dipolar sources, such that we used the inversion technique for dipolar sources described above. Following previous paleomagnetic studies of dusty olivine chondrules, we adopted an experimentally determined value for f’ of 1.87 \citep{fu_solar_2014, lappe_comparison_2013}.

Figures S5 and S6 show the results of the paleointensity experiments. For all samples except DOC3a, we fit for $\Delta NRM$/$\Delta ARM$ using reduced major axis least squares. For DOC3a, we used ordinary least squares because the correlation parameter was $<$0.6 \citep{smith_use_2009}. $\Delta NRM$ was calculated by vector subtraction from the first demagnetization step in the HC component fit, while $\Delta ARM$ was calculated by subtraction from the first acquired ARM step. Table S5 shows a summary of the paleointensities and their quality criteria.

\subsection{Mineralogy}
Mineral compositions (fig. S4 and tables S3 and S4) were analyzed on a JEOL JXA-8200 Superprobe electron probe microanalyzer (EPMA) using WDS in the MIT Electron Microprobe Facility. BSE images were obtained with the same instrument (fig. S4). The EPMA was operated at an accelerating voltage of 15 kV and a beam current of 10 nA, and natural and synthetic standards were used for calibration. The counting times were typically 40 s per element, and the 1$\sigma$ SDs of the accumulated counts were 0.5 to 1.0\% from counting statistics. The raw data were corrected for matrix effects using the CITZAF package \citep{armstrong_citzaf_1995}. To identify the origin of the magnetic signal of these two samples, we mapped the magnetization of chondrules using the QDM at MIT (fig. S4) \citep{fu_high-sensitivity_2020, glenn_micrometer-scale_2017}. The sensor to sample distance was $\sim$5 $\mu$m, and the map resolution was 1.17 $\mu$m. The QDM maps were obtained after the demagnetization of an ARM applied to DOC2 (200 $\mu$T bias with an AF of 145 mT) to 145 mT (fig. S4C) and after the application of an ARM (200 $\mu$T bias with an AF of 100 mT) to DOC6b (fig. S4F).

\bibliography{references}

\begin{thebibliography}{}
\expandafter\ifx\csname natexlab\endcsname\relax\def\natexlab#1{#1}\fi
\providecommand{\url}[1]{\href{#1}{#1}}
\providecommand{\dodoi}[1]{doi:~\href{http://doi.org/#1}{\nolinkurl{#1}}}
\providecommand{\doeprint}[1]{\href{http://ascl.net/#1}{\nolinkurl{http://ascl.net/#1}}}
\providecommand{\doarXiv}[1]{\href{https://arxiv.org/abs/#1}{\nolinkurl{https://arxiv.org/abs/#1}}}

\bibitem[{Alexander {et~al.}(2007)Alexander, Fogel, Yabuta, \&
  Cody}]{alexander_origin_2007}
Alexander, C. M.~O., Fogel, M., Yabuta, H., \& Cody, G.~D. 2007, Geochimica et
  Cosmochimica Acta, 71, 4380, \dodoi{10.1016/j.gca.2007.06.052}

\bibitem[{Alexander {et~al.}(2018)Alexander, Greenwood, Bowden, Gibson, Howard,
  \& Franchi}]{alexander_mutli-technique_2018}
Alexander, C. M.~O., Greenwood, R.~C., Bowden, R., {et~al.} 2018, Geochimica et
  Cosmochimica Acta, 221, 406, \dodoi{10.1016/j.gca.2017.04.021}

\bibitem[{Andrews(2020)}]{andrews_observations_2020}
Andrews, S.~M. 2020, Annual Review of Astronomy and Astrophysics, 58, 483,
  \dodoi{10.1146/annurev-astro-031220-010302}

\bibitem[{Armstrong(1995)}]{armstrong_citzaf_1995}
Armstrong, J.~T. 1995, Microbeam Analysis, 4, 177

\bibitem[{Atreya {et~al.}(2018)Atreya, Crida, Guillot, Lunine, Madhusudhan, \&
  Mousis}]{atreya_weiss_2018}
Atreya, S.~K., Crida, A., Guillot, T., {et~al.} 2018, in Saturn in the 21st
  {Century} (Cambridge: Cambridge University Press), 5--43,
  \dodoi{10.1017/9781316227220.002}

\bibitem[{Bai(2016)}]{bai_toward_2016}
Bai, X. 2016, Astrophys. J., 821, 80

\bibitem[{Bai(2017)}]{bai_global_2017}
Bai, X.-N. 2017, The Astrophysical Journal, 845, 75,
  \dodoi{10.3847/1538-4357/aa7dda}

\bibitem[{Bai \& Goodman(2009)}]{bai_heat_2009}
Bai, X.-N., \& Goodman, J. 2009, The Astrophysical Journal, 701, 737,
  \dodoi{10.1088/0004-637X/701/1/737}

\bibitem[{Bonal {et~al.}(2007)Bonal, Bourot-Denise, Quirico, Montagnac, \&
  Lewin}]{bonal_organic_2007}
Bonal, L., Bourot-Denise, M., Quirico, E., Montagnac, G., \& Lewin, E. 2007,
  Geochimica et Cosmochimica Acta, 71, 1605, \dodoi{10.1016/j.gca.2006.12.014}

\bibitem[{Brasser \& Mojzsis(2020)}]{brasser_partitioning_2020}
Brasser, R., \& Mojzsis, S.~J. 2020, Nature Astronomy, 4, 492,
  \dodoi{10.1038/s41550-019-0978-6}

\bibitem[{Carporzen {et~al.}(2011)Carporzen, Weiss, Elkins-Tanton, Shuster,
  Ebel, \& Gattacceca}]{carporzen_magnetic_2011}
Carporzen, L., Weiss, B.~P., Elkins-Tanton, L.~T., {et~al.} 2011, Proc. Natl.
  Acad. Sci. USA, 108, 6386

\bibitem[{Clarke {et~al.}(2001)Clarke, Gendrin, \&
  Sotomayor}]{clarke_dispersal_2001}
Clarke, C.~J., Gendrin, A., \& Sotomayor, M. 2001, Monthly Notices of the Royal
  Astronomical Society, 328, 485, \dodoi{10.1046/j.1365-8711.2001.04891.x}

\bibitem[{Cournede {et~al.}(2015)Cournede, Gattacceca, Gounelle, Rochette,
  Weiss, \& Zanda}]{cournede_early_2015}
Cournede, C., Gattacceca, J., Gounelle, M., {et~al.} 2015, Earth and Planetary
  Science Letters, 410, 62, \dodoi{10.1016/j.epsl.2014.11.019}

\bibitem[{Davidson {et~al.}(2019)Davidson, Alexander, Stroud, Busemann, \&
  Nittler}]{davidson_mineralogy_2019}
Davidson, J., Alexander, C. M.~O., Stroud, R.~M., Busemann, H., \& Nittler,
  L.~R. 2019, Geochimica et Cosmochimica Acta, 265, 259,
  \dodoi{10.1016/j.gca.2019.08.032}

\bibitem[{DeMeo \& Carry(2014)}]{demeo_solar_2014}
DeMeo, F.~E., \& Carry, B. 2014, Nature, 505, 629, \dodoi{10.1038/nature12908}

\bibitem[{Desch {et~al.}(2018)Desch, Kalyaan, \& Alexander}]{desch_effect_2018}
Desch, S.~J., Kalyaan, A., \& Alexander, C. M.~O. 2018, The Astrophysical
  Journal Supplement Series, 238, 11, \dodoi{10.3847/1538-4365/aad95f}

\bibitem[{Desch {et~al.}(2010)Desch, Morris, Connolly, \&
  Boss}]{desch_critical_2010}
Desch, S.~J., Morris, M.~A., Connolly, H.~C., \& Boss, A.~P. 2010, The
  Astrophysical Journal, 725, 692, \dodoi{10.1088/0004-637X/725/1/692}

\bibitem[{Desch {et~al.}(2012)Desch, Morris, Connolly, \&
  Boss}]{desch_importance_2012}
---. 2012, Meteoritics \& Planetary Science, 47, 1139,
  \dodoi{10.1111/j.1945-5100.2012.01357.x}

\bibitem[{Ebert {et~al.}(2018)Ebert, Render, Brennecka, Burkhardt, Bischoff,
  Gerber, \& Kleine}]{ebert_ti_2018}
Ebert, S., Render, J., Brennecka, G.~A., {et~al.} 2018, Earth and Planetary
  Science Letters, 498, 257, \dodoi{10.1016/j.epsl.2018.06.040}

\bibitem[{Einsle {et~al.}(2016)Einsle, Harrison, Kasama, Conbhuí, Fabian,
  Williams, Woodland, Fu, Weiss, \& Midgley}]{einsle_multi-scale_2016}
Einsle, J.~F., Harrison, R.~J., Kasama, T., {et~al.} 2016, Am. Mineral., 101,
  2070

\bibitem[{Fu {et~al.}(2020{\natexlab{a}})Fu, Kehayias, Weiss, Schrader, Bai, \&
  Simon}]{fu_weak_2020}
Fu, R.~R., Kehayias, P., Weiss, B.~P., {et~al.} 2020{\natexlab{a}}, Journal of
  Geophysical Research: Planets, e2019JE006260,
  \dodoi{https://doi.org/10.1029/2019JE006260}

\bibitem[{Fu {et~al.}(2020{\natexlab{b}})Fu, Lima, Volk, \&
  Trubko}]{fu_high-sensitivity_2020}
Fu, R.~R., Lima, E.~A., Volk, M. W.~R., \& Trubko, R. 2020{\natexlab{b}},
  Geochemistry, Geophysics, Geosystems, 21, e2020GC009147,
  \dodoi{10.1029/2020GC009147}

\bibitem[{Fu {et~al.}(2021)Fu, Volk, Bilardello, Libourel, Lesur, \&
  Dor}]{fu_fine-scale_2021}
Fu, R.~R., Volk, M. W.~R., Bilardello, D., {et~al.} 2021, AGU Advances, 2,
  e2021AV000486, \dodoi{10.1029/2021AV000486}

\bibitem[{Fu {et~al.}(2014)Fu, Weiss, Lima, Harrison, Bai, Desch, Ebel, Suavet,
  Wang, Glenn, Le~Sage, Kasama, Walsworth, \& Kuan}]{fu_solar_2014}
Fu, R.~R., Weiss, B.~P., Lima, E.~A., {et~al.} 2014, Science, 346, 1089

\bibitem[{Glenn {et~al.}(2017)Glenn, Fu, Kehayias, Sage, Lima, Weiss, \&
  Walsworth}]{glenn_micrometer-scale_2017}
Glenn, D.~R., Fu, R.~R., Kehayias, P., {et~al.} 2017, Geochemistry, Geophysics,
  Geosystems, 18, 3254, \dodoi{10.1002/2017GC006946}

\bibitem[{Hartmann {et~al.}(1998)Hartmann, Calvet, Gullbring, \&
  D'Alessio}]{hartmann_accretion_1998}
Hartmann, L., Calvet, N., Gullbring, E., \& D'Alessio, P. 1998, Astrophys. J.,
  495, 385, \dodoi{10.1086/305277}

\bibitem[{Hartmann {et~al.}(2016)Hartmann, Herczeg, \&
  Calvet}]{hartmann_accretion_2016}
Hartmann, L., Herczeg, G., \& Calvet, N. 2016, Annual Review of Astronomy and
  Astrophysics, 54, 135, \dodoi{10.1146/annurev-astro-081915-023347}

\bibitem[{Hertwig {et~al.}(2019)Hertwig, Kimura, Ushikubo, Defouilloy, \&
  Kita}]{hertwig_26al-26mg_2019}
Hertwig, A.~T., Kimura, M., Ushikubo, T., Defouilloy, C., \& Kita, N.~T. 2019,
  Geochimica et Cosmochimica Acta, 253, 111, \dodoi{10.1016/j.gca.2019.02.020}

\bibitem[{Kirschvink(1980)}]{kirschvink_least-squares_1980}
Kirschvink, J.~L. 1980, Geophys. J. R. Astr. Soc., 62, 699

\bibitem[{Kirschvink {et~al.}(2008)Kirschvink, Kopp, Raub, Baumgartner, \&
  Holt}]{kirschvink_rapid_2008}
Kirschvink, J.~L., Kopp, R.~E., Raub, T.~D., Baumgartner, C.~T., \& Holt, J.~W.
  2008, Geochemistry, Geophysics, Geosystems, 9, \dodoi{10.1029/2007GC001856}

\bibitem[{Kita \& Ushikubo(2011)}]{kita_evolution_2011}
Kita, N.~T., \& Ushikubo, T. 2011, Meteorit. Planet. Sci., 47, 1108

\bibitem[{Kleine {et~al.}(2020)Kleine, Budde, Burkhardt, Kruijer, Worsham,
  Morbidelli, \& Nimmo}]{kleine_non-carbonaceouscarbonaceous_2020}
Kleine, T., Budde, G., Burkhardt, C., {et~al.} 2020, Space Science Reviews,
  216, 55, \dodoi{10.1007/s11214-020-00675-w}

\bibitem[{Kruijer {et~al.}(2020)Kruijer, Kleine, \& Borg}]{kruijer_great_2020}
Kruijer, T.~S., Kleine, T., \& Borg, L.~E. 2020, Nature Astronomy, 4, 32,
  \dodoi{10.1038/s41550-019-0959-9}

\bibitem[{Lappe {et~al.}(2013)Lappe, Feinberg, Muxworthy, \&
  Harrison}]{lappe_comparison_2013}
Lappe, S.-C. L.~L., Feinberg, J.~M., Muxworthy, A., \& Harrison, R.~J. 2013,
  Geochemistry, Geophysics, Geosystems, 14, 2143, \dodoi{10.1002/ggge.20141}

\bibitem[{Lappe {et~al.}(2011)Lappe, Church, Kasama, Fanta, Bromiley,
  Dunin‐Borkowski, Feinberg, Russell, \& Harrison}]{lappe_mineral_2011}
Lappe, S.-C. L.~L., Church, N.~S., Kasama, T., {et~al.} 2011, Geochemistry,
  Geophysics, Geosystems, 12, \dodoi{10.1029/2011GC003811}

\bibitem[{Lichtenberg {et~al.}(2021)Lichtenberg, Draazkowska, Schönbächler,
  Golabek, \& Hands}]{lichtenberg_bifurcation_2021}
Lichtenberg, T., Drazkowska, J., Schönbächler, M., Golabek, G.~J., \&
  Hands, T.~O. 2021, Science, 371, 365, \dodoi{10.1126/science.abb3091}

\bibitem[{Lima \& Weiss(2016)}]{lima_ultra-high_2016}
Lima, E.~A., \& Weiss, B.~P. 2016, Geochemistry, Geophysics, Geosystems, 17,
  3754, \dodoi{10.1002/2016GC006487}

\bibitem[{Morbidelli {et~al.}(2015)Morbidelli, Walsh, O'Brien, Minton, \&
  Bottke}]{morbidelli_dynamical_2015}
Morbidelli, A., Walsh, K.~J., O'Brien, D.~P., Minton, D.~A., \& Bottke, W.~F.
  2015, Asteroids IV, 493, \dodoi{10.2458/azu_uapress_9780816532131-ch026}

\bibitem[{Nagashima {et~al.}(2017)Nagashima, Krot, \&
  Komatsu}]{nagashima_26al26mg_2017}
Nagashima, K., Krot, A.~N., \& Komatsu, M. 2017, Geochimica et Cosmochimica
  Acta, 201, 303, \dodoi{10.1016/j.gca.2016.10.030}

\bibitem[{Nagy {et~al.}(2019)Nagy, Williams, Tauxe, \&
  Muxworthy}]{nagy_nano_2019}
Nagy, L., Williams, W., Tauxe, L., \& Muxworthy, A.~R. 2019, Geochemistry,
  Geophysics, Geosystems, 20, 2907, \dodoi{10.1029/2019GC008319}

\bibitem[{Owen {et~al.}(2012)Owen, Clarke, \& Ercolano}]{owen_theory_2012}
Owen, J.~E., Clarke, C.~J., \& Ercolano, B. 2012, Mon. Not. R. Astron. Soc.,
  422, 1880, \dodoi{10.1111/j.1365-2966.2011.20337.x}

\bibitem[{Picogna {et~al.}(2019)Picogna, Ercolano, Owen, \&
  Weber}]{picogna_dispersal_2019}
Picogna, G., Ercolano, B., Owen, J.~E., \& Weber, M.~L. 2019, Monthly Notices
  of the Royal Astronomical Society, 487, 691, \dodoi{10.1093/mnras/stz1166}

\bibitem[{Rosotti {et~al.}(2013)Rosotti, Ercolano, Owen, \&
  Armitage}]{rosotti_interplay_2013}
Rosotti, G.~P., Ercolano, B., Owen, J.~E., \& Armitage, P.~J. 2013, Monthly
  Notices of the Royal Astronomical Society, 430, 1392,
  \dodoi{10.1093/mnras/sts725}

\bibitem[{Schneider {et~al.}(2020)Schneider, Burkhardt, Marrocchi, Brennecka,
  \& Kleine}]{schneider_early_2020}
Schneider, J.~M., Burkhardt, C., Marrocchi, Y., Brennecka, G.~A., \& Kleine, T.
  2020, Earth and Planetary Science Letters, 551, 116585,
  \dodoi{10.1016/j.epsl.2020.116585}

\bibitem[{Schrader {et~al.}(2017)Schrader, Nagashima, Krot, Ogliore, Yin,
  Amelin, Stirling, \& Kaltenbach}]{schrader_distribution_2017}
Schrader, D.~L., Nagashima, K., Krot, A.~N., {et~al.} 2017, Geochim. Cosmochim.
  Acta, 201, 275

\bibitem[{Scott {et~al.}(1992)Scott, Keil, \& Stöffler}]{scott_shock_1992}
Scott, E. R.~D., Keil, K., \& Stöffler, D. 1992, Geochimica et Cosmochimica
  Acta, 56, 4281, \dodoi{10.1016/0016-7037(92)90268-N}

\bibitem[{Scott \& Krot(2013)}]{scott_chondrites_2013}
Scott, E. R.~D., \& Krot, A.~N. 2013, in Planets, {Asteroids}, {Comets}, and
  the {Solar} {System}, Vol.~2, Treatise on {Geochemistry}, ed. H.~D. Holland
  \& K.~K. Turekian (Amsterdam: Elsevier Science), 66--137

\bibitem[{Scott {et~al.}(2018)Scott, Krot, \& Sanders}]{scott_isotopic_2018}
Scott, E. R.~D., Krot, A.~N., \& Sanders, I.~S. 2018, The Astrophysical
  Journal, 854, 164, \dodoi{10.3847/1538-4357/aaa5a5}

\bibitem[{Selkin {et~al.}(2000)Selkin, Gee, Tauxe, Meurer, \&
  Newell}]{selkin_effect_2000}
Selkin, P.~A., Gee, J.~S., Tauxe, L., Meurer, W.~P., \& Newell, A.~J. 2000,
  Earth Planet. Sci. Lett., 183, 403

\bibitem[{Shah {et~al.}(2018)Shah, Williams, Almeida, Nagy, Muxworthy, Kovács,
  Valdez-Grijalva, Fabian, Russell, Genge, \&
  Dunin-Borkowski}]{shah_oldest_2018}
Shah, J., Williams, W., Almeida, T.~P., {et~al.} 2018, Nature Communications,
  9, 1, \dodoi{10.1038/s41467-018-03613-1}

\bibitem[{Siron {et~al.}(2021{\natexlab{a}})Siron, Fukuda, Kimura, \&
  Kita}]{siron_new_2021}
Siron, G., Fukuda, K., Kimura, M., \& Kita, N.~T. 2021{\natexlab{a}},
  Geochimica et Cosmochimica Acta, 293, 103, \dodoi{10.1016/j.gca.2020.10.025}

\bibitem[{Siron {et~al.}(2021{\natexlab{b}})Siron, Kita, Fukuda, \&
  Kimura}]{siron_high_2021}
Siron, G., Kita, N.~T., Fukuda, K., \& Kimura, M. 2021{\natexlab{b}}, Lunar and
  Planetary Science Conference, 1639.
\newblock \url{https://ui.adsabs.harvard.edu/abs/2021LPI....52.1639S/abstract}

\bibitem[{Smith(2009)}]{smith_use_2009}
Smith, R.~J. 2009, American Journal of Physical Anthropology, 140, 476,
  \dodoi{10.1002/ajpa.21090}

\bibitem[{Stephenson(1993)}]{stephenson_three-axis_1993}
Stephenson, A. 1993, Journal of Geophysical Research: Solid Earth, 98, 373,
  \dodoi{10.1029/92JB01849}

\bibitem[{Sutton {et~al.}(2017)Sutton, Alexander, Bryant, Lanzirotti, Newville,
  \& Cloutis}]{sutton_bulk_2017}
Sutton, S., Alexander, C. M.~O., Bryant, A., {et~al.} 2017, Geochimica et
  Cosmochimica Acta, 211, 115, \dodoi{10.1016/j.gca.2017.05.021}

\bibitem[{Suzuki {et~al.}(2016)Suzuki, Ogihara, Morbidelli, Crida, \&
  Guillot}]{suzuki_evolution_2016}
Suzuki, T.~K., Ogihara, M., Morbidelli, A., Crida, A., \& Guillot, T. 2016,
  Astronomy \& Astrophysics, 596, A74, \dodoi{10.1051/0004-6361/201628955}

\bibitem[{Tauxe \& Staudigel(2004)}]{tauxe_strength_2004}
Tauxe, L., \& Staudigel, H. 2004, Geochemistry, Geophysics, Geosystems, 5,
  \dodoi{10.1029/2003GC000635}

\bibitem[{Testi {et~al.}(2014)Testi, Birnstiel, Ricci, Andrews, Blum,
  Carpenter, Dominik, Isella, Natta, Williams, \& Wilner}]{testi_dust_2014}
Testi, L., Birnstiel, T., Ricci, L., {et~al.} 2014, Protostars and Planets VI,
  339, \dodoi{10.2458/azu_uapress_9780816531240-ch015}

\bibitem[{Tikoo {et~al.}(2014)Tikoo, Weiss, Cassata, Shuster, Gattacceca, Lima,
  Suavet, Nimmo, \& Fuller}]{tikoo_decline_2014}
Tikoo, S.~M., Weiss, B.~P., Cassata, W.~S., {et~al.} 2014, Earth and Planetary
  Science Letters, 404, 89, \dodoi{10.1016/j.epsl.2014.07.010}

\bibitem[{Turner {et~al.}(2014)Turner, Fromang, Gammie, Klahr, Lesur, Wardle,
  \& Bai}]{turner_transport_2014}
Turner, N.~J., Fromang, S., Gammie, C., {et~al.} 2014, in Protostars and
  {Planets} {VI} (University of Arizona Press), 411--432

\bibitem[{Wang {et~al.}(2017)Wang, Weiss, Bai, Downey, Wang, Wang, Suavet, Fu,
  \& Zucolotto}]{wang_lifetime_2017}
Wang, H., Weiss, B.~P., Bai, X.-N., {et~al.} 2017, Science, 355, 623,
  \dodoi{10.1126/science.aaf5043}

\bibitem[{Wang {et~al.}(2019)Wang, Bai, \& Goodman}]{wang_global_2019}
Wang, L., Bai, X.-N., \& Goodman, J. 2019, The Astrophysical Journal, 874, 90,
  \dodoi{10.3847/1538-4357/ab06fd}

\bibitem[{Watson(1956)}]{watson_test_1956}
Watson, G.~S. 1956, Geophysical Journal International, 7, 160,
  \dodoi{10.1111/j.1365-246X.1956.tb05561.x}

\bibitem[{Weiss {et~al.}(2021)Weiss, Bai, \& Fu}]{weiss_history_2021}
Weiss, B.~P., Bai, X.-N., \& Fu, R.~R. 2021, Science Advances, 7, eaba5967,
  \dodoi{10.1126/sciadv.aba5967}

\bibitem[{Weiss {et~al.}(2007)Weiss, Lima, Fong, \&
  Baudenbacher}]{weiss_paleomagnetic_2007}
Weiss, B.~P., Lima, E.~A., Fong, L.~E., \& Baudenbacher, F.~J. 2007, J.
  Geophys. Res.: Solid Earth, 112, B09105, doi:10.1029/2007JB004940

\end{thebibliography}
\bibliographystyle{aasjournal}

\section*{Acknowledgments}
We thank K. Righter and the Meteorite Working Group for allocating the samples, R. Fu and N. Kita for helpful discussions, and R. Walsworth and R. Fu for sharing quantum diamond microscopy technology and techniques.This study was supported by NASA Discovery program (grant NNM16AA09C) and by the NSF (grant DMS-1521765). Conceptualization: C.S.B. and B.P.W. Methodology: C.S.B., B.P.W., and E.A.L. Investigation: C.S.B. and B.P.W. Interpretation: C.S.B., B.P.W., J.F.J.B., and X.-N.B. Visualization: C.S.B. Supervision: B.P.W. Writing—initial drafts: C.S.B. and B.P.W. Writing—review and editing: B.P.W., J.F.J.B., X.-N.B., E.A.L., N.C., and E.N.M. Funding acquisition: B.P.W. The authors declare that they have no competing interests. All data needed to evaluate the conclusions in the paper are present in the paper and available through the Magnetics Information Consortium (MagIC) Database (earthref.org/MagIC/17129).

\nocite{ebert_ti_2018, schneider_early_2020, demeo_solar_2014, morbidelli_dynamical_2015, sutton_bulk_2017, testi_dust_2014, desch_effect_2018, selkin_effect_2000, siron_new_2021, siron_high_2021, hertwig_26al-26mg_2019, nagashima_26al26mg_2017}

\section*{Supplementary materials}
Supplementary text and materials are available at: \href{https://www.science.org/doi/10.1126/sciadv.abj6928}{DOI:10.1126/sciadv.abj6928}.

\end{document}